\documentclass[prd,twocolumn,showpacs,preprintnumbers,amsmath,amssymb]{revtex4}
\usepackage{graphicx}
\usepackage{dcolumn}
\usepackage{bm}

\newcommand{\psip}{\psi(2S)}
\newcommand{\psipp}{\psi(3770)}
\newcommand{\jpsi}{J/\psi}
\newcommand{\DDbar}{D\overline{D}}
\newcommand{\EE}{e^+e^-}
\newcommand{\MM}{\mu^+\mu^-}

\newcommand{\pp}{\pi^+\pi^-}
\newcommand{\ppb}{p\bar{p}}

\newcommand{\ppjpsi}{\pi^+\pi^- J/\psi}
\newcommand{\ra}{\rightarrow}

\newcommand{\rhopi}{\rho\pi}
\newcommand{\omegapi}{\omega\pi^0}
\newcommand{\beq}{\begin{equation}}
\newcommand{\eeq}{\end{equation}}
\newcommand{\bfg}{\begin{figure}}
\newcommand{\efg}{\end{figure}}
\newcommand{\bitm}{\begin{itemize}}
\newcommand{\eitm}{\end{itemize}}
\newcommand{\bnum}{\begin{enumerate}}
\newcommand{\enum}{\end{enumerate}}
\newcommand{\btbl}{\begin{table}}
\newcommand{\etbl}{\end{table}}
\newcommand{\btbu}{\begin{tabular}}
\newcommand{\etbu}{\end{tabular}}
\newcommand{\aggg}{a_{3g}}
\newcommand{\agcc}{a_{\gamma}}
\newcommand{\agee}{a_c}
\newcommand{\rdr}{\sigma_{exp}}
\newcommand{\rpdr}{\sigma^{\prime}_{exp}}

\begin{document}


\title{The Interference between Continuum and Resonance in $\EE \rightarrow
c\overline{c}$  Experiment} 

\author{P.Wang$^{1}$}
 \email{wangp@mail.ihep.ac.cn}
\author{C.Z.Yuan$^{1}$}
\author{X.H.Mo$^{1,2}$}
\author{D.H.Zhang$^{1}$}

\affiliation{
$^1$Institute of High Energy Physics, CAS, Beijing 100039, China\\
$^2$China Center of Advanced Science and Technology (World Laboratory),
Beijing 100080, China}

\date{\today}
	   
\begin{abstract}
$\EE$ experiments at charmonium production threshold are reviewed, it is
found that the contribution of the continuum process via virtual photon
has been neglected in almost all the experiments and most channels
analyzed. It is pointed out that the contribution of the continuum 
part may affect the final results significantly 
in $\psip$ and $\psipp$ decays, while the interference between the
continuum amplitude and the resonance amplitude
may even affect the $\jpsi$ decays as well as the $\psip$ and $\psipp$.
This  leads to the revise of the analysis of 
strong and electromagnetic amplitude in $\psip$ decays, including 
$\psip \rightarrow VP$ which is the long lasting puzzle 
between $\jpsi$ and $\psip$ decays. For $\psipp$ physics, 
a large constructive interference
for light hadron modes and destructive interference for $\DDbar$
could be responsible for the discrepancy between 
the larger cross section of inclusive hadrons by direct measurement of
$\EE \rightarrow \psipp \rightarrow hadrons$
than the $\DDbar$ cross section measured 
using $D$ single-tag and double-tag method.
\end{abstract}
\maketitle

\section{Introduction}

It is well known that the $\EE$ experiments have lots of advantages in particle
physics study: large cross section, small background, and well-determined
initial state (both four-momentum and quantum numbers).
There were lots of $\EE$ experiments, there are still many $\EE$ experiments
working, and there will be further experiments 
build to continue the experimental
study. With the energy ranges from $\pp$ threshold up to TeV scale,
these experiments contribute a lot to the knowledge of the world around us. 
Among them,
there are some working at the $\tau$-charm energy region, 
where the $J^{PC}=1^{--}$
charmonium states are produced and studied, including Mark-I, 
Mark-II, Mark-III, DM2, Crystal Ball, BES and so on. 
Recently, CLEO working at CESR decided lower its energy from the
$B\overline{B}$ 
threshold to the charm threshold~\cite{cleoc}, and BES working
at BEPC decided upgrade both the accelerator and the detector 
to make a factory-like experiment~\cite{bes3}, these two 
experiments will reach an extremely high precision in the
study of the charmed mesons and the dynamics of the charmonium 
states decay in this energy range.

$\jpsi$, the first vector charmonium state discovered in
1974~\cite{jpsi}, gains
lots of attention due to its surprising narrow width and strong coupling 
to $\EE$ state. Since then, it has been used as an ideal laboratory for light 
hadron spectroscopy and charmonium decay dynamics study, which are the 
essential tasks of the low energy QCD. The attempt of figuring out the strong 
decays of $\jpsi$ via three-gluon and electromagnetic decays via 
one-photon annihilation reveals the relative phase between these two 
amplitudes may be large, with the help of some pure electromagnetic decay 
modes of $\jpsi$ like $\omega\pi^0$~\cite{phase}. 
This is an important information since 
it indicates there would be no interference between
these two amplitudes. This situation will be further studied in this paper.

$\psip$, the radially excited spin triplet state 
of $\jpsi$, has also the narrow
nature and strong coupling to $\EE$ state, but most 
impressing feature found in the $\psip$ study is the 
abnormal suppression of some decay modes compared with the
corresponding $\jpsi$ decays based on perturbative QCD predictions. 
This suppression was first observed by the Mark-II experiment in 
vector pseudoscalar (VP) decay modes like $\rho\pi$ and 
$K^*\overline{K}$~\cite{mk2}, and confirmed by BES~\cite{besVP}
(referred as ``$\rho\pi$ puzzle'' in literatures).
Moreover, BES also observed the suppression in 
vector tensor (VT) decays of $\psip$~\cite{besVT}. 
This has led to substantial theoretical efforts in 
solving the problem~\cite{rhopitheo,phase2}, unfortunately, most
of the models were ruled out by the experiments, while some 
others need further experimental data to test.

The $\psipp$, currently regarded as the D wave charmonium state, lies above the
$\DDbar$ threshold, as a consequence, the OZI allowed decays of $\psipp \ra
\DDbar$ would dominate its decays. This picture has been considered
true for pretty long period of time, until recently, a careful study of the
old analyses indicates the $\DDbar$ cross section may be lower than inclusive
hadron cross section of $\psipp$, or in other words, there are substantial
non-$\DDbar$ decays of $\psipp$ state~\cite{nonddbar}.

The three topics in charmonium studies (relative phase between 
strong and electromagnetic decays, ``$\rhopi$ puzzle'', and
non-$\DDbar$ decays of $\psipp$) play important roles 
in understanding the charmonium decay dynamics. In following 
sections, we will examine carefully the
experimental observables and theoretical expectations in 
charmonium study in $\EE$ experiments, to provide a possibility 
of investigating these problems in a self-consistent way by 
considering the unavoidable background process in
$\EE$ experiment, namely, the continuum process. 
We argue that, for any exclusive decay final states
of these charmonia decay, in some cases, the contribution of this 
process to the amplitude may be very important, while in
some other cases, although the direct contribution is relatively 
small, the interference between this term and other dominant 
amplitudes may contribute a non-negligible part, 
which maybe provide a guideline to solve
the three existing problems in the charmonium decays.

The purpose of our paper focuses on locating the natural source of
the existing problems rather than to offer a detailed solutions of them.
So we begin with the inclusive hadronic process to exhibit the experimental
effect on theoretical cross section, then simple assumption is often 
adopted to estimate the function on exclusive processes from the continuum
contribution. At the last of the paper, we also studied the experimental 
condition dependence of the results in case the interference 
was not considered, which is true for most of the existing 
experimental results on $\jpsi$, $\psip$ and $\psipp$ decays.

\section{Experimentally observed cross sections}

We know that $\jpsi$ and $\psip$ decay into light hadrons through 
two interactions: the three-gluon strong interaction and
the one-photon electromagnetic interaction. 
There is in general a relative phase between these two amplitudes. 
This is also true for $\psipp$ in its OZI suppressed decay 
into light hadrons. These two amplitudes and the phase between them
are extracted from experimental data by several authors
for $\jpsi$ and $\psip$~\cite{phase,phase2,phase3}. 
If we denote the amplitude of three-gluon by $\aggg$ and 
one-photon by $\agcc$, both of them can be complex,
the decay rate
\begin{equation}
\label{r}
   \sigma \propto | \aggg + \agcc |^2~~~.
\end{equation}

\begin{figure}[hbt]
\begin{minipage}{3.5cm}
\includegraphics[width=3.5cm,height=2cm]{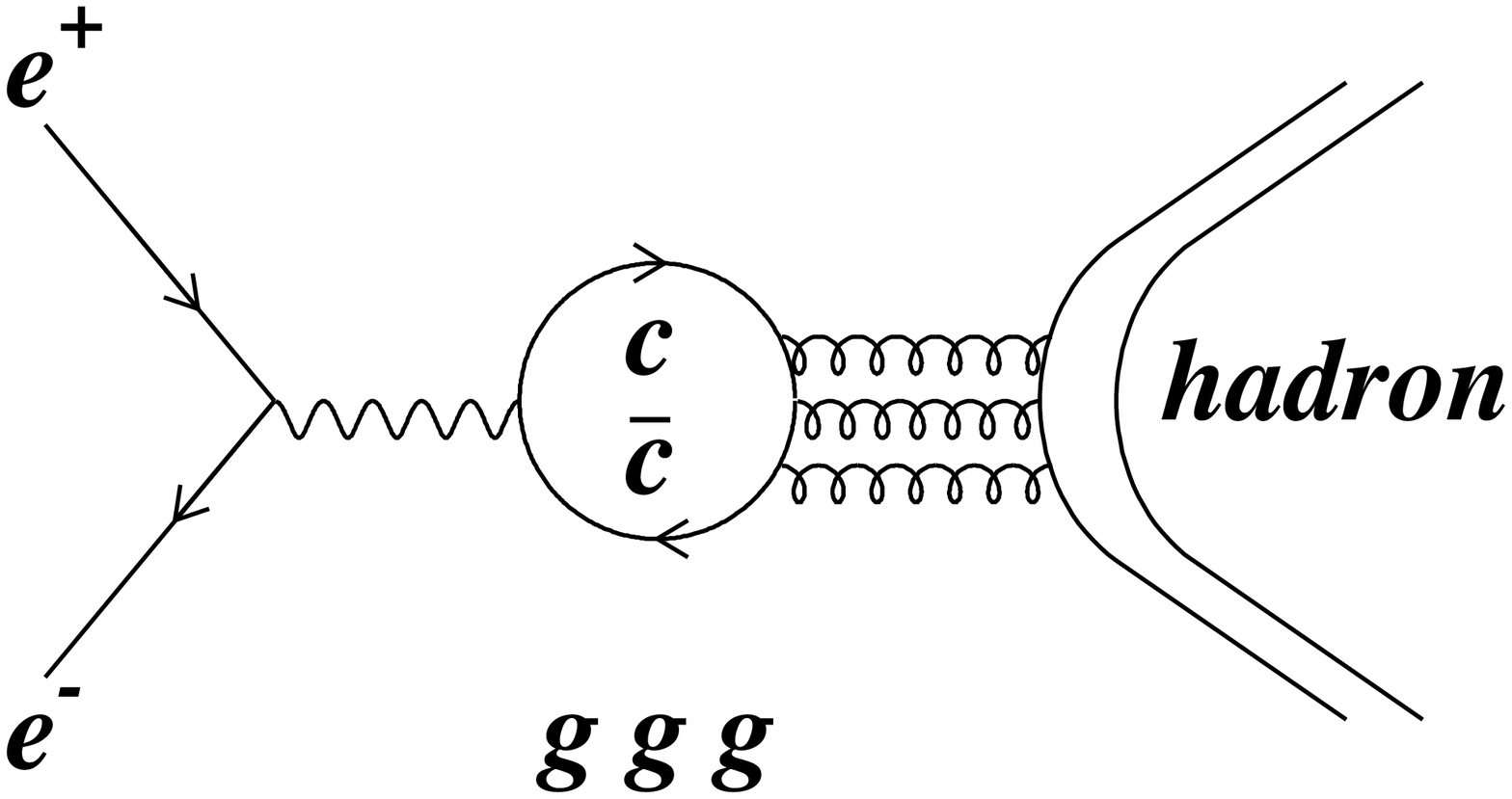}
\center{(a) three gluon process}
\end{minipage}
\hskip 0.5cm
\begin{minipage}{3.5cm}
\includegraphics[width=3.5cm,height=2cm]{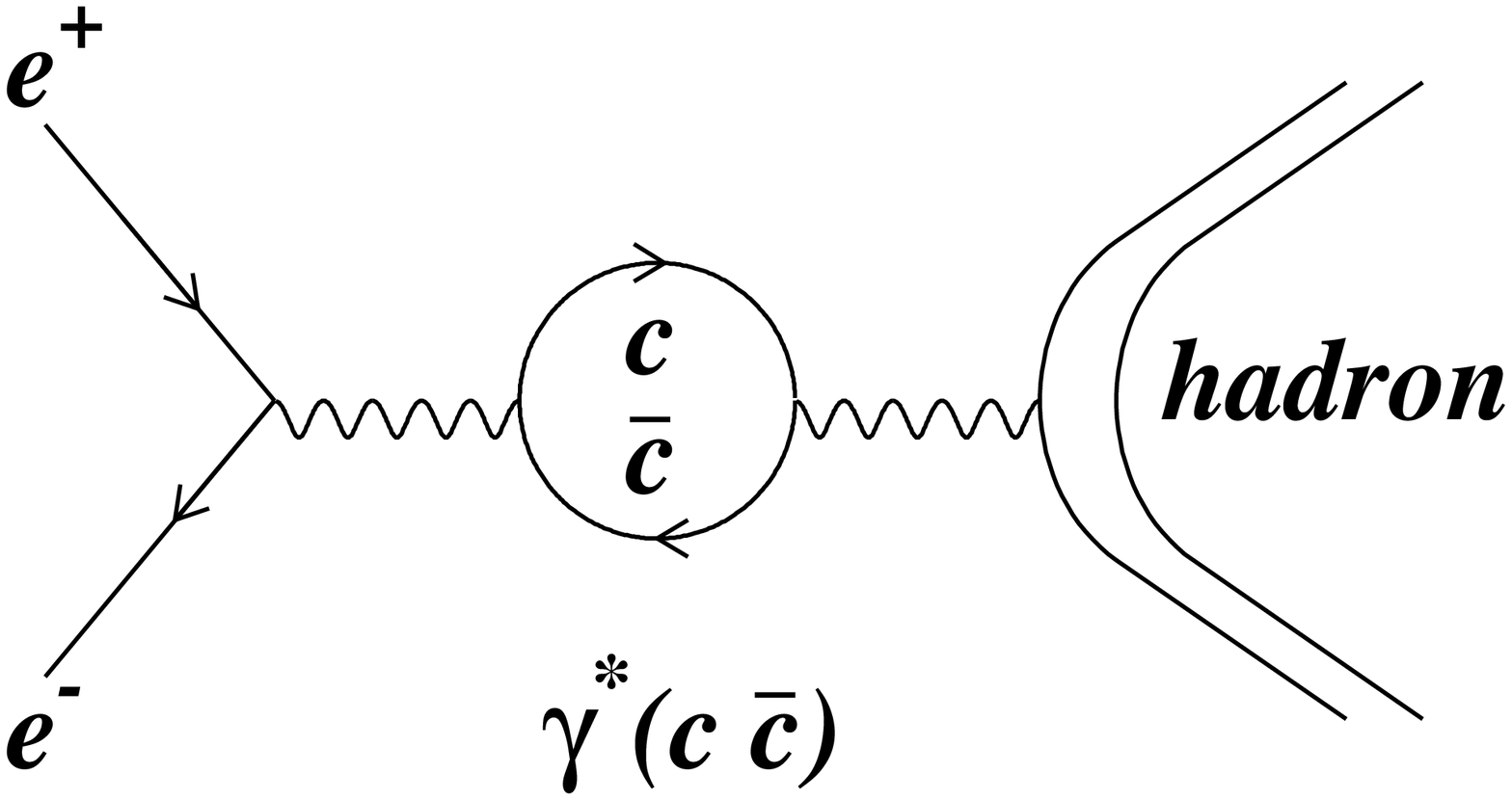}
\center{(b) one photon process}
\end{minipage}
\vskip 0.5cm
\begin{minipage}{5.5cm}
\includegraphics[width=3.5cm,height=2cm]{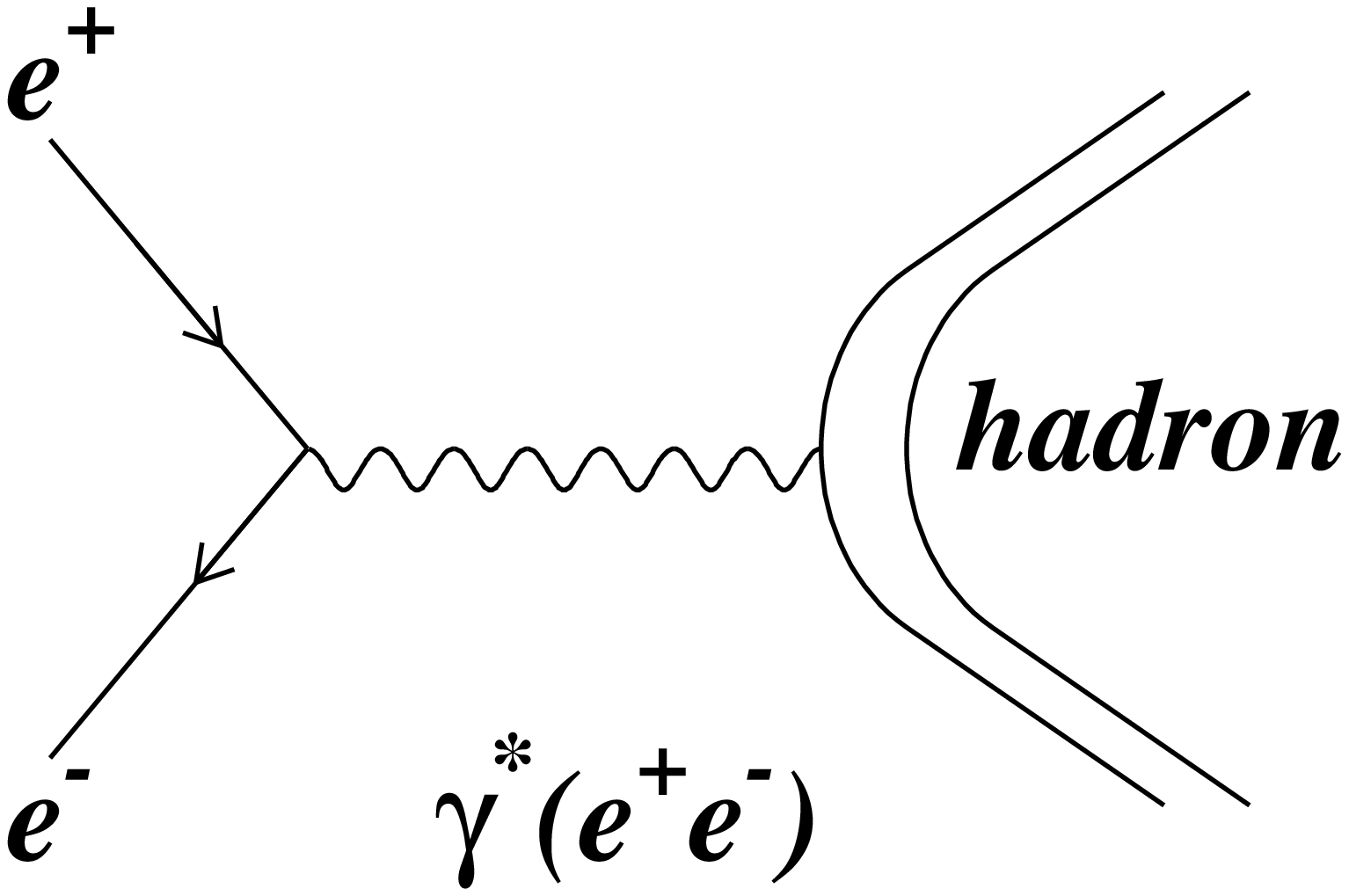}
\center{(c) one photon continuum process}
\end{minipage}
\caption{\label{threefymn} Three Feynman diagrams of hadron process.}
\end{figure}

In $e^+e^-$ colliding beam experiments, the charmonium are produced
from $e^+e^-$ annihilation, there is inevitable another amplitude
\begin{equation}
e^+e^- \rightarrow \gamma^* \rightarrow hadrons
\end{equation}
accompanied with the production of the resonances. This amplitude
does not go through the resonance, but
in general  it could produce the same final hadronic states
as charmonium decays do.  
So there are three  Feynman diagrams corresponding to the
experimentally measured cross sections, i.e.
the three-gluon decays, the one-photon decays, and the 
one-photon continuum process, as illustrated in Fig.~\ref{threefymn}.
The former two amplitudes are associated with the resonance, 
while the last one is a slowly varying function of 
C.M. energy ($\sqrt{s}$). 
To analyze the experimental results, we must take into 
account three amplitudes and two phases.
Taking the amplitude of one-photon continuum as $\agee$, 
the experimentally observed  cross section
\begin{equation}
   \sigma^{\prime} \propto | \aggg + \agcc  + \agee |^2~~~.
\label{rprime}
\end{equation}
In this paper, for simplicity, we define
\begin{equation}
\agee = \frac{e^2{\cal F}(s)}{s} e^{i \phi^{\prime}}~~~,
\end{equation}
where $e$ is the electromagnetic coupling constant, 
${\cal F}(s)$ is the form factor, and 
\begin{equation}
\aggg = \frac{{\cal C}_{ggg}}{s-M^2+i M \Gamma}~~~,
\end{equation}
\begin{equation}
\agcc=\frac{{\cal C}_{\gamma} e^{i\phi}}{s-M^2+i M \Gamma}~~~,
\end{equation}
where ${\cal C}_{ggg}$ and ${\cal C}_{\gamma}$ are taken to be real,
and $M$ and $\Gamma$ are the mass and the width of the resonance.
We shall use this form for general discussions and 
numerical calculations. 

The experimentally observed cross sections in $\EE$ collision
are modified by the initial state radiation. 
For the narrow resonances $\jpsi$ and $\psip$, the observed
cross sections are also distorted by the 
energy spread of the collider. The radiative corrected cross
section is expressed as~\cite{rad.1}

\begin{equation}
\sigma_{r.c.} (\sqrt{s})=\int \limits_{0}^{x_m} dx
F(x,s) \frac{\sigma_{Born}(s(1-x))}{|1-\Pi (s(1-x))|^2}~~~,
\label{radsec}
\end{equation}
where $\sigma_{Born}$ is the Born order cross section.
In the upper limit of the integration $x_m=1-s_m/s$, 
$\sqrt{s_m}$ is the experimentally required minimum invariant
mass of the final state $f$ after losing energy to multi-photon emission;
$F(x,s)$ has been calculated in several references \cite{rad.1, rad.2, rad.3}
and $\Pi (s(1-x))$ is the vacuum polarization factor.

The $e^+e^-$ colliders have finite energy resolution.
The energy resolution function $G(\sqrt{s},\sqrt{s^{\prime}})$ 
is usually described by a Gaussian distribution :
\begin{equation}
G(\sqrt{s},\sqrt{s^{\prime}})=\frac{1}{\sqrt{2 \pi} \Delta}
          e^{ -\frac{(\sqrt{s}-\sqrt{s^{\prime}})^2}{2 {\Delta}^2} },
\end{equation}
where $\Delta$, a function of the beam energy, is the 
C.M. energy spread of the accelerator.
So the experimentally measured resonance cross section,
$\rdr$, is the radiative corrected cross section $\sigma_{r.c.}$
folded with the energy resolution function:
\begin{equation}
\sigma_{exp} (\sqrt{s})=\int \limits_{0}^{\infty} \sigma_{r.c.} 
(\sqrt{s^{\prime}}) G(\sqrt{s^{\prime}},\sqrt{s}) d\sqrt{s^{\prime}}~~~.
\label{expsec}
\end{equation}
Through out this paper, all the physical quantities to be discussed
are experimentally observed ones and the radiative 
correction as well as collider energy spread are implicitly taken 
into account.

In principle, any experiment measures $\rpdr$ should 
subtract the contribution of the continuum part to get the 
physical quantity $\sigma_{exp}$, where
$\sigma^{\prime}_{exp}$ and $\sigma_{exp}$ indicate the experimental
cross sections calculated from Eq.(\ref{expsec}) with the substitution
of $\sigma^{\prime}$ and $\sigma$ from Eq.(\ref{rprime}) and Eq.(\ref{r})
respectively to $\sigma_{B}$ in Eq.(\ref{radsec}).
Unfortunately, up to now, most of the experiments
just neglect this contribution and $\rdr=\rpdr$ is 
assumed for almost all the channels studied, at least at
$J/\psi$ and $\psip$.
The noticeable exceptions are $\Gamma_{ee}$, $\Gamma_{\mu\mu}$ and 
the total width. These quantities are measured, together with the 
resonance mass, by scanning around the peak of the
resonance and then fitting the measured curves with the 
theoretical cross sections. In the fitting, the theoretical 
cross sections always include a continuum term\cite{jpsipscan}. 

The difference between $\rdr$ and $\rpdr$ implies a plausible 
paraphrase in high energy physics literatures.
On one hand, the theoretical analyses are based on $\rdr$, on
the other hand, the experiments actually measure $\rpdr$. 
However, even in the case that 
the continuum amplitude is relatively small, 
such as in $\psip$, certain values of the phase possibly lead to 
non-negligible interference.
For $\psipp$ scan experiment, the inclusive continuum hadron cross 
section is larger than the resonance peak, possible interference 
may contribute a substantial part of the observed cross section.

We now display the effect from the 
continuum amplitude and corresponding phase for $\jpsi$, 
$\psip$ and $\psipp$. To do this, we calculate
the ratio
\beq
k \equiv \frac{ \rpdr - \rdr }{ \rpdr }~~~.
\label{ratiok}
\eeq

In order to see the effect of the relative phase, the
magnitude of $\aggg$, $\agcc$ and $\agee$ are treated as input.
In principle, $\aggg$, $\agcc$ and $\agee$ are different for 
different exclusive modes both in absolute value and in the
relative strength. For illustrative purpose,
following assumption is used for any of the exclusive mode:
the squared moduli of  $\aggg$ and $\agcc$ are
proportional  to their branching ratios
of inclusive hadrons $B({\cal R} \rightarrow ggg \rightarrow hadron)$
and $B({\cal R} \rightarrow \gamma^* \rightarrow hadron)$
given by PDG~\cite{pdg}, and the squared modulus of $\agee$ is assumed to be 
proportional to the Born order $\mu^+\mu^-$ cross section 
multiplied by $R_{had}$ which indicates the hadronic cross
section of the continuum process, and is estimated by 
pQCD~\cite{pdg}.
Table \ref{estimation} lists these inputs for $\jpsi$, $\psip$
and $\psipp$.

\btbl
\caption{\label{estimation} Amplitude estimation for three charmonium
states~\cite{pdg}.
$\sigma_{{\cal R}}$ is total cross section for resonance ${\cal R}$,
(${\cal R} = \jpsi, \psip$, and $\psipp$).}
\vskip 0.2 cm
\center
\btbu{c||ccc} \hline \hline
             &   $\jpsi$    
                        &   $\psip$   
                                  & $\psipp$ \\ \hline \hline        
$| \aggg |^2$& 60\% $\sigma_{\jpsi}$
                        & 15\% $\sigma_{\psip}$
			          & $\sim$ 1\% $\sigma_{\psipp}$ \\
$| \agcc |^2$& 17\% $\sigma_{\jpsi}$
                        & 2.9\% $\sigma_{\psip}$
                                  & $3\times 10^{-5} \sigma_{\psipp}$ \\
$| \agee |^2$& 20~nb
                        & 15~nb
                                 & 13~nb \\ \hline \hline
\etbu
\etbl

\begin{figure}[hbt]
\begin{minipage}{7.5cm}
\includegraphics[width=7.5cm,height=5cm]{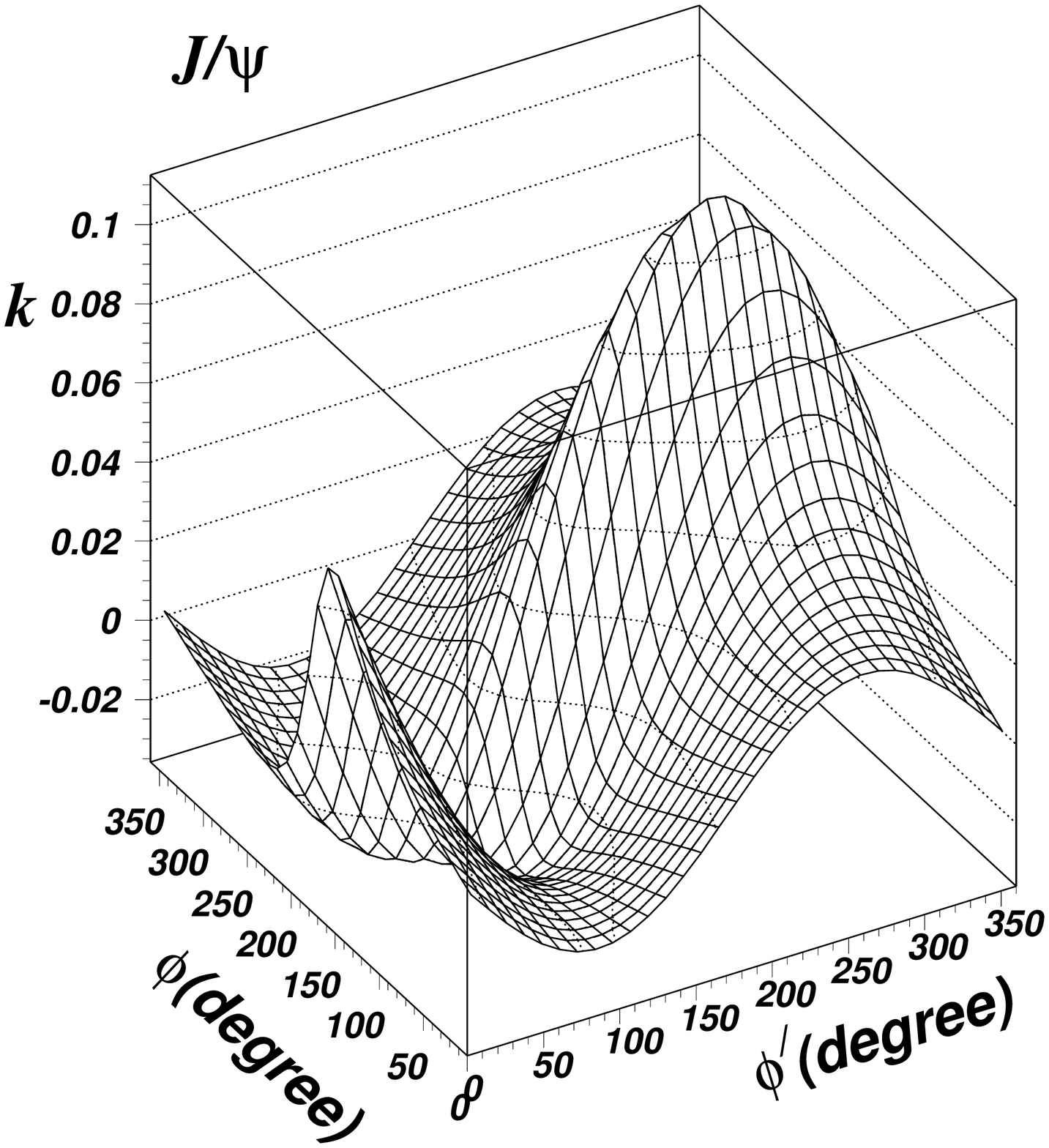}
\center{(a) $k$ for $\jpsi$}
\end{minipage}
\hskip 0.5cm
\begin{minipage}{7.5cm}
\includegraphics[width=7.5cm,height=5cm]{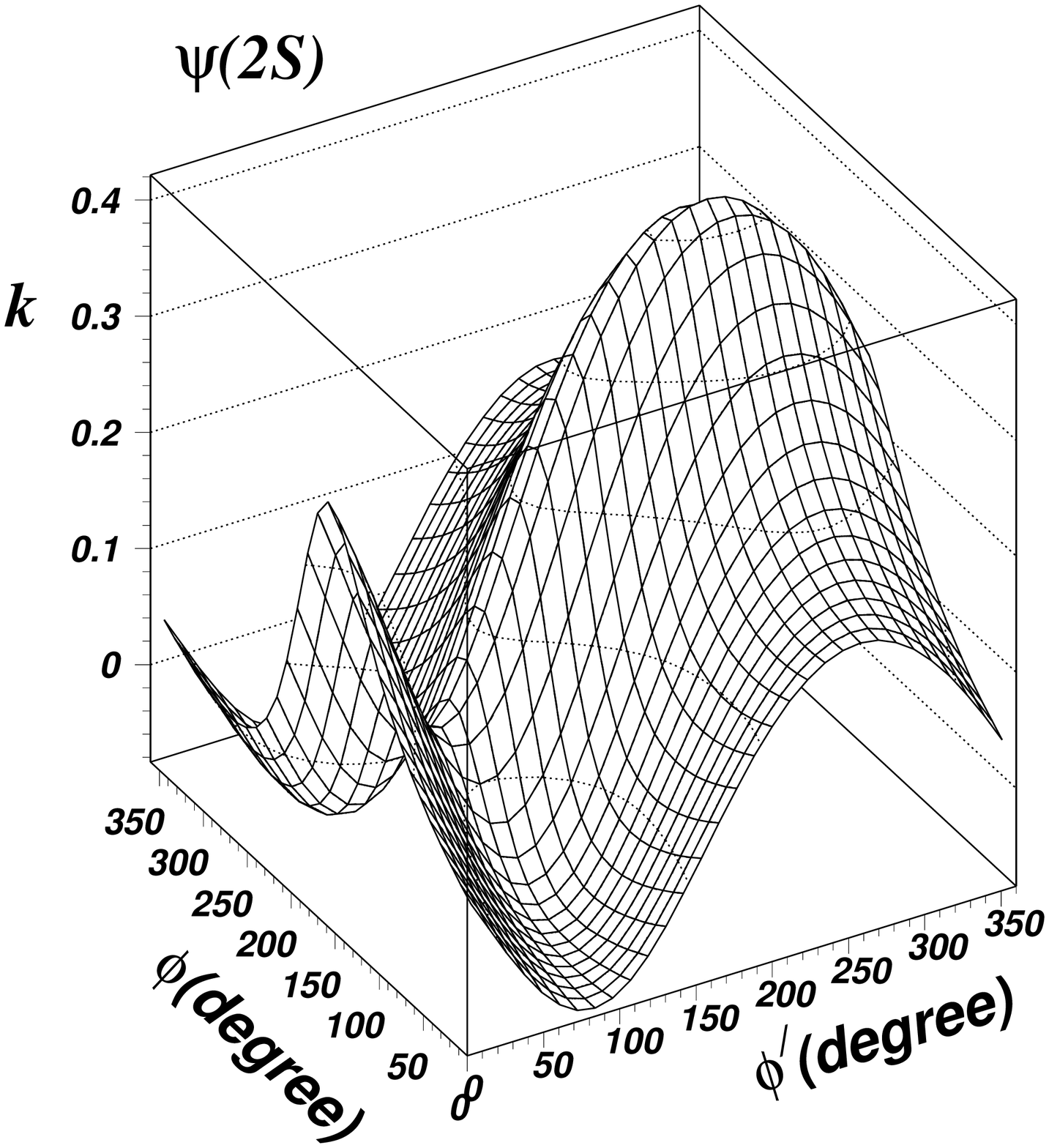}
\center{(b) $k$ for $\psip$}
\end{minipage}
\caption{\label{kratio} $k$ as a function of  $\phi$ and $\phi^{\prime}$,
(a) for $\jpsi$ and (b) for $\psip$. }
\end{figure}

If we use the BEPC energy spread listed by PDG~\cite{pdg},
$\sigma_{\jpsi} \simeq $~3100~nb,
 $\sigma_{\psip} \simeq $~700~nb,
and $\sigma_{\psipp} \simeq $~8 nb are got. Combining with equation 
(\ref{r}), (\ref{rprime}) and (\ref{ratiok}), we could obtain
$k$ as a function of $\phi$ and $\phi^{\prime}$, whose variation 
is shown in Fig.~\ref{kratio}.
It can be seen that for certain values of the two phases,
$k$ could deviates from 0, or equivalently the ratio $\rpdr/ \rdr$ deviates
from 1, which implies that the continuum process may produce non-negligible 
effect in experimental measurement.

\section{$\agee$ in $\EE \rightarrow c \bar{c}$ processes}

\subsection{$\psipp$}

The $\psipp$ is a wide resonance with 
$\Gamma_{tot}=(23.6\pm2.5)$~MeV~\cite{pdg}.
The collider energy spread can be neglected as long as
the standard deviation of the energy resolution function $\Delta$ 
is comparatively small, such as $\Delta \leq 2$~MeV. 
Its maximum cross section of inclusive hadrons in $\EE$ collision  is
8~nb,  while the continuum cross section is about
13~nb. The resonance predominantly decays into $\DDbar$,
while the continuum cross section mainly goes to light hadrons.
The decay rate of $\psipp$ to light hadrons via three-gluon annihilation
even though small in the total width,
e.g. at the order of 1\%, or partial width of
about 230~keV, which means that $|\aggg| \approx 0.08|a_c|$
($\psipp$ decays to light hadrons via one-photon annihilation
is three orders of magnitude lower than the three-gluon
annihilation given above \cite{pdg}, and could be neglected), if it
has interference with continuum amplitude, it
could bring an interference of maximum 1.9~nb in the observed
cross section to light hadrons. Another interference between
the tail of $\psip \rightarrow \aggg$ and $\agee$ at $\sqrt{s}=m_{\psipp}$
brings an interference of maximum 1~nb, while the interference 
between $\psip$ and $\psipp$ at $\sqrt{s}=m_{\psipp}$ is small, 
with 0.07~nb at maximum. The continuum cross section of 
$\EE\rightarrow\DDbar$, in the simple naive quark model, is 
estimated by 
\beq
\sigma_c(s) = \frac{4}{3} \cdot \sigma_{\MM}(s)
\cdot \frac{2P_{D}}{E_{cm}},
\eeq
with $P_{D}$ being the momentum of the $D$ or $\overline{D}$.
After taken into account of radiative correction, it is 
0.019nb. But the interference between $a_c$ and $a_{D}$ is
0.79nb at maximum. Here $a_{D}$ denotes the OZI allowed 
strong interaction amplitude which 
is responsible for $\psipp\rightarrow\DDbar$ decays. 

A possible large constructive interference for light hadrons and, 
at the same time, a large
destructive interference for $D\overline{D}$, could be responsible for 
the larger cross section of inclusive hadrons by direct measurement of
$\EE \rightarrow \psipp \rightarrow hadrons$~\cite{psi''xection}
than the $D\overline{D}$ cross section measured by Mark III Collaboration 
using $D$ single-tag and double-tag~\cite{MARK3}.

As to the exclusive decays, it could make some of the decay modes
with small branching ratios more observable at the resonance. 
For example, if the missing decay modes of $\psip$ like $\rho\pi$ do
appear in $\psipp$ decays, with an enhancement factor~\cite{Rosner}, 
their on-resonance cross section
could be substantially larger than off-resonance
in $\EE$ experiment. Quantitatively, if
${\cal B}(\psipp\rightarrow \rhopi) \approx 4\times10^{-4}$ 
(or equivalently, $\sigma_{\psipp\rightarrow \rho\pi} \approx 0.003$~nb)
as suggested in Ref.~\cite{Rosner}, and
$\sigma(\EE\rightarrow\rhopi)\approx 0.014$~nb
at Born order by the model of Ref.~\cite{Achasov}$^*$
~\footnotetext[1]{The same calculation
gives $\rho\pi$ cross section at
the mass of $\psip$ to be 0.015~nb, which is just below the
current upper limit of the branching ratio $2.8\times 10^{-5}$
or upper limit of the cross section 0.02~nb by BES~\cite{besVP}.},
then the maximum interference could be 0.011~nb,
much larger than the pure contribution from $\psipp \ra \rho\pi$. 
Comparing the cross sections on and off $\psipp$ peak,
$\psipp \ra \rho\pi$ could be seen through the interference with the
continuum amplitude.

\subsection{$\psip$}

As can be seen in Fig.~\ref{kratio}~(b), the ratio $\rpdr/\rdr$ 
could deviate from $1$ substantially. 
For each exclusive decay channels, $k$ 
could be different, due to the magnitudes of $\agcc$ and $\aggg$,
have different coefficients and $\agee$, if estimated by form factors, 
are of different functions of the energy. 
This must be taken into account in the fitting of $\agcc$, $\aggg$ 
and the phase in between.
In general, with the interference between $\agee$ and the resonance, 
the maximum height of each exclusive channel does not
necessarily coincide with the maximum height of the inclusive hadrons
on which data are taken. We shall
take the $\MM$ channel as an example in the next section.

In $\psip$ final state analyses, it is noticeable that the observed 
cross sections of some electromagnetic processes, such as
$\psip \rightarrow \pp$, $\psip\rightarrow \omegapi$,
and the famous puzzling process $\psip\rightarrow \rhopi$,
are three to four orders of magnitude smaller than 
the inclusive hadron cross section of the continuum process, which is 
about 15~nb. Form factor estimation \cite{fofa1}
gives these cross sections comparable to the magnitudes 
off the resonance\cite{wym}. 
It implies that a substantial part of the
experimentally measured cross section could comes from the continuum
amplitude $\agee$ instead of the $\psip$ decays, and interference
between these two amplitudes may even affect the measured quantities
further.  Therefore it is essential to know the production rate
of $\pp$, $\omegapi$ and $\rhopi$ due to the continuum process
in order to get their correct branching ratios of the 
$\psip$ decays. 

In order to know whether the observed suppression of VP and VT modes 
in $\psip$ decays are due to the absence of strong interaction
amplitude, or the destructive interference between the electromagnetic
and the strong amplitudes, or just an incidental destructive interference 
between these two and the continuum process in particular experiment, 
the amplitude $\agee$ must be taken into account. 

\subsection{$\jpsi$}

From Fig.~\ref{kratio}~(a), it is seen that the interference between the 
amplitude $\agee$ and the resonance is at the order of a few percent 
which is much smaller than that of $\psip$. It is also smaller than 
the statistical and systematic uncertainties of current measurements. 
Nevertheless, for future high precision measurements such as the
proposed CLEO-c~\cite{cleoc} and BES-III~\cite{bes3}, when the accuracy 
goes to a few per mille level, it should be taken into account. 

\section{The dependence on experimental conditions}

In this section, we discuss the dependence of the 
observed cross section in $\EE$ collision on the experimental
conditions. The most crucial experiment conditions are the 
accelerator energy spread and the beam energy setting.
The former will smear the intrinsic width of the resonance 
so that change the relative contribution between the
resonance and the continuum, while the latter will affect
the relative contribution as well as the absolute
correction to the total rate due to the interference.
The invariant mass cut or equivalent requirement in data 
analysis will also affect the relative contribution of 
resonance and continuum due to the different energy dependence
of the cross sections in radiative correction.

\subsection{Dependence on collider energy resolution}

\begin{figure}[hbt]
\begin{center}
\includegraphics[width=8.5 cm,height=7.cm]{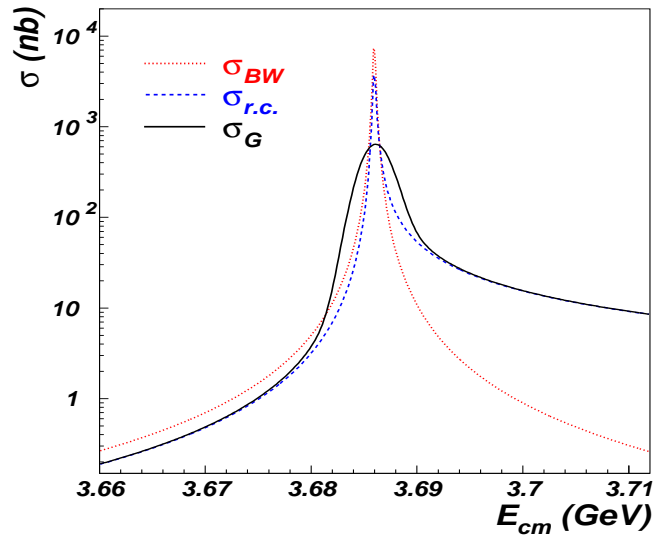}
\caption{\label{cmpbrg} Total cross section at $\psip$:
$\sigma_{BW}$ for Breit-Wigner cross section,
$\sigma_{r.c.}$ the cross section with radiative correction,
and $\sigma_{exp}$ the measured cross section on a collider with
$\Delta=1.3$ MeV.}
\end{center}
\end{figure}

For narrow resonance like $\jpsi$ and $\psip$, with intrinsic
widths much narrower than the energy resolution of the 
current $\EE$ colliders, we don't observe their original resonance 
curve. Instead, what we actually measure is the 
resonance smeared by the finite  energy resolution of the collider.
In Fig.~\ref{cmpbrg}, three cross sections are depicted: the
Breit-Wigner cross section, the cross section after
radiative correction (Eq.(\ref{radsec})), and the experimentally 
measured cross section (Eq.(\ref{expsec})). 

\begin{figure}[hbt]
\begin{center}
\includegraphics[width=7.5cm,height=8.cm]{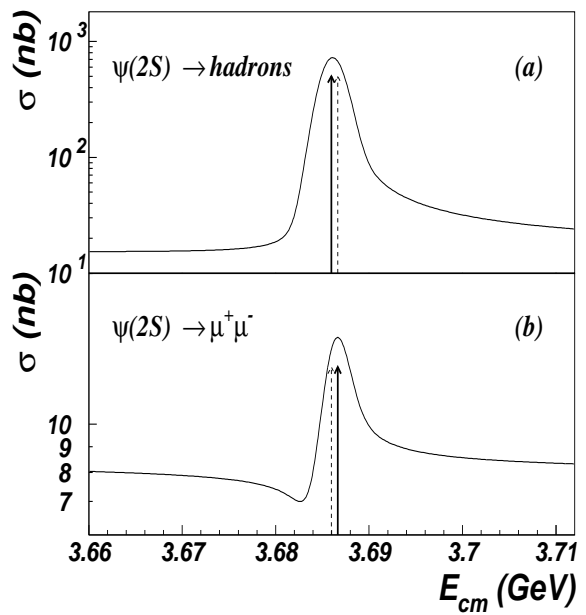}
\caption{\label{cmphup} Cross sections in the vicinity of $\psip$
for hadron (a) and $\MM$ (b) final states.
The solid line with arrow indicates the peak position
and the dashed line with arrow the shift of the other peak position.}
\end{center}
\end{figure}

In actual experiments, data are naturally taken at the energy which 
yields the maximum height of the inclusive hadrons. This 
energy is not the nominal mass of the resonance but somewhat higher, 
neither does it coincide with the maximum height of each exclusive 
channels due to the interference effect with $\agee$. For comparison, 
Fig.~\ref{cmphup}, depicts the observed cross sections of
inclusive hadrons and $\mu^+\mu^-$ pairs. Two arrows in the figure
denote the different positions of the maximum heights of the cross 
sections. It is well know that the radiative correction 
reduces the height of the peak and shift the maximum height of 
the resonance peak upwards, and 
the energy resolution of the collider both reduces the height of 
the peak and shifts it more profoundly. 
Such shift, depends on the energy resolution of the collider,
in general could be different for inclusive hadrons and for 
each exclusive channels. For example, the peak of $\mu^+\mu^-$ curve 
is shifted more than that of the inclusive hadrons, 
to 0.81~MeV above the $\psip$ nominal mass for BEPC energy 
spread, which is 1.3~MeV.

It is clear that the shape of the 
observed cross section is much different from that of Breit-Wigner.
However, the energy smear hardly affects the continuum part of the 
cross section. So in the observed cross section, what
proportion comes from the contribution of continuum 
and  interference is sensitive to the energy spread.
The larger the collider energy spread is, the more share 
the continuum part contributes in the observed cross section. 

\subsection{Dependence on beam energy}

\begin{figure}[hbt]
\begin{center}
\includegraphics[width=7.5cm,height=8.cm]{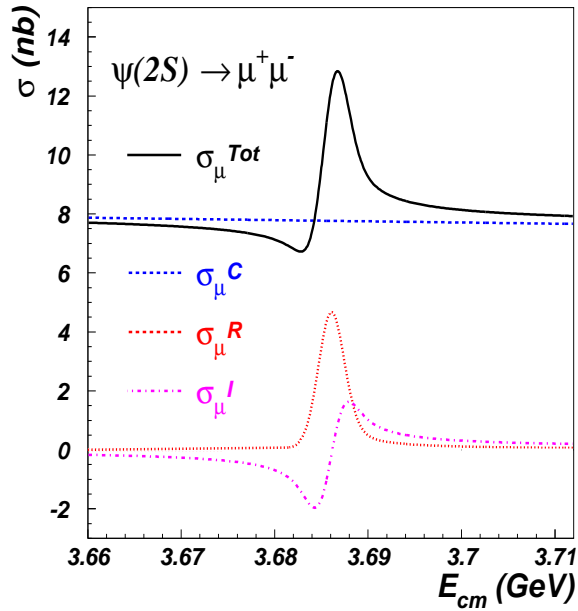}
\caption{\label{cmpuirc} Contributions of three parts to the cross section:
dashed line for QED continuum ($\sigma^C$);
dotted line for resonance ($\sigma^R$);
dash dotted line for interference($\sigma^I$);
solid line for total cross section($\sigma^{Tot}$).}
\end{center}
\end{figure}

For demonstration of such dependence, we show the curve of $\MM$ 
channel, for its dynamics is clear and there is no unknown 
parameter. It is similar to those hadronic channels 
in $\psip$ decays which only go through electromagnetic interaction, 
such as $\omegapi$ and $\pi^+\pi^-$.
Since this is an exclusive channel, there is interference between
the continuum and the $\psip$ amplitudes.
Such interference can be seen clearly from the scan of the $\psip$,
as shown in Fig.~\ref{cmpuirc}. 

The relative contribution of the resonance and the continuum
changes rapidly as the energy changes, and the interference term 
between these two amplitudes also varies with energy. 
The latter could change from negative to positive as the 
energy passes across the nominal mass of the resonance.
In the actual experimental situation, it is important to know 
the beam energy precisely, which is needed in the delicate task
to subtract the contribution of $\agee$. Possible uncertainty and 
drift of the beam energy need to be taken into account in 
the determination of the systematic errors.    

\subsection{Dependence on invariant mass requirement}

The magnitude of the continuum part of the cross section is sensitive 
to the upper limit of the integration in the calculation of the 
radiative correction, i.e., it depends sensitively on the 
invariant mass cut in the experiment (Eq.~(\ref{radsec})). This is 
because that the Born order cross section of the continuum term goes 
up as C.M. energy decreases. If the event selection uses a very loose
cut, then the cross section from continuum part could be 
very large. This is particularly true for exclusive channels, 
because QCD predicts the form factors to be powers of $1/Q^2$. 
While the resonance part is not sensitive to the invariant
mass cut, since the Breit-Wigner formula serves as a natural cut by 
itself. Mathematically, in the upper limit of the integration of
Eq.~(\ref{radsec}), as long as $(1-s_m/s) \gg \Gamma/M$, the integration
is not sensitive to the upper limit, where $\Gamma$ and 
$M$ are the total width and mass of the resonance respectively.

So in the observed cross section, what proportion comes from 
the contribution of continuum and interference is sensitive 
to the events selection criteria. Many of a time, it is not the
invariant mass cut directly applied to the data, instead it is
affected by many cuts, like the momentum cut, kinematic fit,
collinearity cut and so on. In this case, how much is the 
contribution of the continuum and the interference could only
be calculated by Monte Carlo simulation. Qualitatively,
looser invariant mass requirement in event selection would
increase the share of the continuum part of the contribution.

\bigskip

It is worth noting here that in principle if $\agee$ is not considered
correctly, different experiments will give different results 
to the same quantity, like the exclusive branching ratio
of the resonance, due to the dependence on beam energy spread,
beam energy setting, and invariant mass requirement in event
selection. This point is especially important for the time being, 
since the beam spreads for different accelerators are 
much different, and events selection criteria
is very different because of the big background in the channels
analyzed~\cite{jpsirep}.

\section{Summary and Perspectives}

As we have emphasized in the foregoing discussion, the amplitude $\agee$, 
by itself or through interference with the resonance, 
could contribute significantly to the observed cross sections
in $\EE$ experiments on charmonium physics.
Its treatment depends sensitively on the experimental details,
this has not been fully addressed in both
$\EE$ experiments and theoretical analyses based on these results.
So far, most of the measurements are crude, with large statistical
and systematic uncertainties, so this problem has been outside of 
the purview of concern. Now with large $\jpsi$ and $\psip$ samples 
from BES-II~\cite{bes2data} and forthcoming high precision 
experiments CLEO-c~\cite{cleoc} and BES-III~\cite{bes3}, 
the effect of $\agee$ needs to be addressed properly. 

To study the continuum contribution, the most promising 
way is to do energy scan for every exclusive mode 
in the vicinity of the resonance, so that
both the amplitudes and the relative phases could be fit out
simultaneously. In case this is not practicable, data sample
off the resonance with comparable integrated luminosity
as on the resonance should be collected to measure $|\agee|$,
which could give an estimation of its contribution in the
decay modes studied. The theoretical analyses based on
current available $\EE$ data, particularly on $\psip$ may need
to be revised correspondingly. 

In fact, another way to free ourselves from the effect of 
the continuum is to analyze the decay product from higher
energy experiments. For example, $\jpsi$ decays could be
measured from $\psip \rightarrow \ppjpsi$ ,
$\psip$ could be studied from $\ppb$ annihilation experiments or 
from $B$ decays at the $B$ factories.

\acknowledgments
The authors wish to thank their colleagues of the BES collaboration
for discussions. This work is supported in part by
the National Natural Science Foundation of China under contracts 
No. 19991483 and 100 Talents Program of the Chinese Academy of Sciences
under contract No. U-25.

\end{document}